# A SCALAR FIELD AND THE EINSTEIN VACUUM IN MODERN KALUZA-KLEIN THEORY

Paul S. Wesson[1] and James M. Overduin[2,3]

[1]Department of Physics and Astronomy, University of Waterloo, Waterloo, ON, N2L 3G1, Canada.

[2]Department of Physics, Astronomy and Geosciences, Towson University, Towson, MD, 21252, U.S.A.

[3]Department of Physics and Astronomy, Johns Hopkins University, 3400 N. Charles St., Baltimore, MD, 21218, U.S.A.

Abstract: Five-dimensional relativity as an extension of general relativity has field equations that simplify considerably given the adoption of a new gauge. The result is a scalar field governed by the Klein-Gordon equation, in an empty spacetime with a cosmological constant governed by Einstein's equations. The main application is to the properties of massive particles in a vacuum-dominated universe.

Note: Changes from v.1: New last paragraph, Appendix

Key Words: Higher dimensions, Scalar field, Vacuum, Cosmological constant

# A SCALAR FIELD AND THE EINSTEIN VACUUM
# IN MODERN KALUZA-KLEIN THEORY

1. Introduction

In the search for a unified field theory, much attention has been paid to the five-dimensional analog of the four-dimensional equations of general relativity. The former equations contain the latter by virtue of Campbell's embedding theorem. However, the 5D equations are difficult to solve. Hence the importance of gauges, which are essentially 5D frameworks for 4D solutions of various types. There are two gauges in widespread use, namely the canonical one of Space-Time-Matter theory and the warp one of Membrane theory [1, 2]. Both help in understanding the nature of 4D vacuum energy, whose density is measured by the cosmological constant. A new gauge will be given below, which describes the classical vacuum governed by the empty Einstein equations, plus a scalar field governed by an equation of Klein-Gordon type.

Gauges are essential to any higher-dimensional form of general relativity. Algebraically, these constrain the metric so as to make the field equations tractable. Physically, they focus attention on the properties of the sources for the fields. In the basic extension of general relativity from 4 to 5 dimensions, the two gauges in current use both deal with the properties of the vacuum. In Space-Time-Matter theory, the aim is to give a geometrical description of matter using a non-compactified 5D metric with a prefactor on the 4D part which is quadratic in the extra coordinate, and guarantees the embedding in 5D of any solution of the 4D Einstein equations which is empty of ordinary matter but has vacuum energy density measured by the cosmological constant. In Membrane theory, the aim is to concentrate the interactions of particles along a singular hypersurface or membrane in 5D, using a metric with a prefactor on the 4D part which is exponential in the extra coordinate, and relates the masses of particles to the properties of the vacuum. Both of these gauges have large literatures [1, 2]. For Space-Time-Matter theory, it is possible to deal with non-vacuum, ordinary matter by allowing the metric coefficients to depend on the extra coordinate [3]. However, the (pure) canonical gauge deals only with vacuum, as does the warp metric of Membrane theory. Neither of these gauges deals with the scalar field, which is represented by the extra diagonal component of the 5D metric, and is characteristic of theories of Kaluza-Klein type.

In what follows, this deficiency will be resolved by the introduction of a third gauge. This has some properties in common with the old ones, but has the important advantage of including the scalar field in addition to the vacuum. The scalar field obeys the standard Klein-Gordon equation, while the vacuum is any solution of the empty Einstein equations. The present account is short, but there are implications for particles in vacuum which should repay future investigation.

The notation is standard, with units chosen so that $G$, $h$ and $c$ are all unity, except where they are made explicit to aid understanding.

2.  A Gauge With Vacuum and a Scalar Field

The 5D coordinates are taken to be $x^A = x^\gamma, l$ where $x^\gamma = 0,123$ refer to time and space, while $x^4 = l$ refers to the extra dimension. The field equations for 5D relativity are commonly taken in terms of the Ricci tensor as

$$R_{AB} = 0 \quad (A, B = 0,123,4) \quad . \tag{1}$$

To solve these equations, the 5D metric can be simplified by using 4 of the 5 available degrees of coordinate freedom to set the potentials of electromagnetic type to zero, via $g_{4\alpha} = 0$ [1, 2]. The remaining degree of coordinate freedom is used in the canonical and warp gauges to set the scalar field to a constant, via $|g_{44}| = 1$. But to bring out the effects of this field, it can be specified by $g_{44} \equiv \varepsilon \Phi^2$, where $\Phi = \Phi(x^\alpha, l)$ and $\varepsilon = \pm 1$ allows for both a spacelike and timelike extra dimension. (The extra dimension for $\varepsilon = +1$ need not have the physical nature of a time so there is no problem with closed historical paths.) The 5D line element then takes the form

$$dS^2 = g_{\alpha\beta}(x^\gamma, \ell) dx^\alpha dx^\beta + \varepsilon \Phi^2(x^\gamma, l) dl^2 \quad . \tag{2}$$

This includes the 4D line element $ds^2 \equiv g_{\alpha\beta} dx^\alpha dx^\beta$, and allows the use of the 4D proper time $s$ as parameter in the study of dynamics (see below). With metric (2), the field equations (1) can be conveniently grouped into sets of 10 (tensor), 4 (vector) and 1 (scalar), thus:

$$G_{\alpha\beta} = 8\pi T_{\alpha\beta}$$

$$8\pi T_{\alpha\beta} \equiv \frac{\Phi_{,\alpha;\beta}}{\Phi} - \frac{\varepsilon}{2\Phi^2} \left\{ \frac{\Phi_{,4} g_{\alpha\beta,4}}{\Phi} - g_{\alpha\beta,44} + g^{\lambda\mu} g_{\alpha\lambda,4} g_{\beta\mu,4} \right.$$

$$\left. - \frac{g^{\mu\nu} g_{\mu\nu,4} g_{\alpha\beta,4}}{2} + \frac{g_{\alpha\beta}}{4} \left[ g^{\mu\nu}_{,4} g_{\mu\nu,4} + \left( g^{\mu\nu} g_{\mu\nu,4} \right)^2 \right] \right\} \quad . \tag{3}$$

$$P^{\beta}_{\alpha;\beta} = 0$$

$$P^{\beta}_{\alpha} \equiv \frac{1}{2\Phi} \left( g^{\beta\sigma} g_{\sigma\alpha,4} - \delta^{\beta}_{\alpha} g^{\mu\nu} g_{\mu\nu,4} \right) \quad . \tag{4}$$

$$\Box \Phi = -\frac{\varepsilon}{2\Phi} \left[ \frac{g^{\lambda\beta}_{,4} g_{\lambda\beta,4}}{2} + g^{\lambda\beta} g_{\lambda\beta,44} - \frac{\Phi_{,4} g^{\lambda\beta} g_{\lambda\beta,4}}{\Phi} \right]$$

$$\Box \Phi \equiv g^{\alpha\beta} \Phi_{,\alpha;\beta} \quad . \tag{5}$$

Here a comma denotes the partial derivative, and a semicolon denotes the standard (4D) covariant derivative. These equations are algebraically general, and can be applied to any physical problem where gravitational and scalar fields are dominant.

To focus on the vacuum and scalar field in spacetime, the metric coefficients in (2) can be restricted by writing

$$g_{\alpha\beta}(x^{\gamma}, l) = \exp(l\Phi/L) \, \bar{g}_{\alpha\beta}(x^{\gamma}), \quad \Phi = \Phi(x^{\gamma}) \quad . \tag{6}$$

Here $L$ is a constant length whose meaning will shortly become clear. The gauge (6), when substituted into the field equations (3) – (5), yields some interesting results.

The effective energy-momentum tensor (3) is given by

$$8\pi T_{\alpha\beta} = \frac{\Phi_{,\alpha;\beta}}{\Phi} - \frac{\varepsilon}{2L^2} g_{\alpha\beta} \qquad (7.1)$$

The second term here is typical of a 4D Einstein space with cosmological constant $\Lambda = -\varepsilon/2L^2$. This is positive for a spacelike extra dimension $(\varepsilon = -1)$ and negative for a timelike extra dimension $(\varepsilon = +1)$. The trace of (7.1) is

$$8\pi T = \frac{\Box \Phi}{\Phi} - \frac{2\varepsilon}{L^2} \quad , \qquad (7.2)$$

which will be used below.

The tensor (4) and its associated scalar are given by

$$P_\alpha^\beta = \frac{-3}{2L} \delta_\alpha^\beta \qquad (8.1)$$

$$P = \frac{-6}{L} \quad . \qquad (8.2)$$

Since the 4-tensor $P_\alpha^\beta$ is conserved, $P$ is obviously a constant for the spacetime.

The scalar equation (5) can be written

$$\Box \Phi + \frac{\varepsilon}{L^2}\Phi = 0 \quad . \tag{9.1}$$

This relation has many well-known applications in physics. For example, when $\Phi$ depends only on the spatial coordinates, it is the Helmholtz equation if $\varepsilon = +1$, the diffusion equation if $\varepsilon = -1$, and the Schrodinger equation if the coefficient of the second term is the kinetic energy of a nonrelativistic particle. When $\Phi$ depends on all four spacetime coordinates, (9.1) is formally identical for $\varepsilon = +1$ with the Klein-Gordon equation. The latter is the wave equation for a relativistic particle of rest mass $m$, where $L$ is then the Compton wavelength of the particle $h/mc$. It should be noted that for the 5D canonical gauge, it is possible to recover the wave function if the extra dimension is timelike, when the extra component of the geodesic equation yields the Klein-Gordon equation [4]. For the new gauge (6), the real or complex nature of the scalar field $\Phi$ is not specified, and its physical interpretation depends partly on $\varepsilon = \pm 1$. (In this regard, $l \to il$ also formally changes the sign of the last term in the 5D metric.) It is due to the fact that $\varepsilon$ can in principle have either sign that it is kept explicit in the present account. This choice affects not only the interpretation of the scalar field equation (9.1) but also the sign of the 4D scalar curvature. This is given in general and for gauge (6) by

$$R = \frac{\varepsilon}{4\Phi^2}\left[ g^{\mu\nu}{}_{,4}\, g_{\mu\nu,4} + \left(g^{\mu\nu} g_{\mu\nu,4}\right)^2 \right] = \frac{3\varepsilon}{L^2} \quad . \tag{9.2}$$

This may be compared to (7.2), which with (9.1) gives the trace of the Einstein tensor as $G = -R$, as expected.

In the above, the 5D field equations (1) have been written for the general metric (2) in the form (3)-(5), and the application of the gauge (6) has been seen to lead to the remarkably simple relations (7)-(9). To complete the formal part of the analysis, it is natural to ask about the paths of test particles, or geodesics. It has been known for some time that 5D *null* paths with $dS^2 = 0$ correspond to 4D paths of photons with $ds^2 = 0$ *and* massive particles with $ds^2 > 0$ [4, 5]. To illustrate this in the present context, set $dS^2 = 0$ in (2) as constrained by (6), choose $\varepsilon = -1$ for a spacelike extra dimension, and specify appropriately a fiducial value of the proper time. The result is

$$l = \frac{2L}{\Phi} \log\left(\frac{2L}{s}\right) \quad . \tag{10}$$

It is seen that the 'size' of the fifth dimension slowly decreases as proper time increases.

3.  Discussion and Conclusion

Kaluza-Klein theory in its general form is a unified account of gravitational, electromagnetic and scalar fields, whose potentials are represented by the $g_{\alpha\beta}$, $g_{4\alpha}$ and $g_{44}$ components of the 5D metric tensor. The 5D field equations (1) are expected to contain the Einstein field equations, the Maxwell equations and the Klein-Gordon equation. However, because the theory is covariant, these fields may be mixed, and appear in guises dependent on the choice of coordinates or gauge. In the present account, the $g_{4\alpha}$ potentials were eliminated by the choice of the starting metric (2) so as to concentrate on the effects of the gravitational and scalar fields. However, for the gauge (6) the tensor $P_\alpha^\beta$ of (4) is still finite, as are the energy-momentum tensor $T_{\alpha\beta}$ of (3) and the scalar potential $\Phi$ of (5). To make a physical interpretation of the algebraic results (7)-(9), the following comments may be useful.

The energy-momentum tensor (7.1) has two components, of which one is associated with a cosmological constant $\Lambda = -\varepsilon/2L^2$, as mentioned before. Here $L$ is the length scale introduced to the gauge (6) for the consistency of physical dimensions, and so gets a physical meaning. If the coupling constant is gravitational, it is common to interpret $\Lambda$ as a measure of the energy density and pressure of the vacuum, via $\rho_v = \Lambda c^4/8\pi G = -p_v$. However, this is not strictly necessary, as $\Lambda$ has natural dimensions of (length)$^{-2}$ and already matches the dimensions of the Einstein tensor. The same comment applies to the other contribution to $T_{\alpha\beta}$, which is the curvature or energy density due to the scalar field $\Phi$. It should be noted that $\Phi(x^\gamma)$ here is a field which is not found in standard general relativity or 4D scalar-tensor theories [6]. It could be classical or quantum in origin, and real or complex in nature (see below). By the trace $T$ of (7.2), the magnitudes of the $\Phi$ and $\Lambda$ contributions may have the same or opposite signs, depending on whether $\varepsilon = -1$ for a spacelike extra dimension or $\varepsilon = +1$ for a timelike extra dimension.

The scalar-field equation (9.1) is well known in physics, and has several interpretations as outlined before. However, it comes from the $R_{44} = 0$ component of the field equations (1), and therefore lies outside standard general relativity. The solutions of (9.1) depend crucially on the choice of $\varepsilon = \pm 1$. In canonical-type 5D metrics, the dynamics of a test particle is known to be monotonic ($\varepsilon = -1$) or oscillatory ($\varepsilon = +1$), with quantum characteristics in the latter case [5]. For the present metric, it is therefore feasible to choose $\varepsilon = +1$ and take $\Phi(x^\gamma)$ to be complex, in which case (9.1) is simply the Klein-Gordon equation of old wave mechanics. The length $L$ is then related to the rest mass $m$ of the particle associated with the $\Phi$-waves: $L = h/mc$. The introduction of Planck's constant here is reasonable, since the gravitational and scalar sectors of the 5D theory may involve different coupling constants. Indeed, quantum couplings in semi-classical relativity have been considered before [7]. Alternatively, in view of previous comments about the energy-momentum tensor, (9.1) may be interpreted in terms of perturbations in the vacuum.

The tensor $P_\alpha^\beta$ of (8.1) and its associated scalar $P$ of (8.2) are the most difficult of the objects in 5D relativity to interpret in a physical sense. One of the reasons for this is that the physical dimensions of these quantities is (length)$^{-1}$, not the (length)$^{-2}$ typical of the gravitational and scalar sectors. Because of this, some workers in Membrane theory have sought to 'square' $P_\alpha^\beta$ and add this to the gravitational energy-momentum tensor as the contribution of the stress in the membrane typical of that theory [1]. However, while it is possible to express $P_\alpha^\beta$ in terms of the extrinsic (Gaussian) curvature of an $l-$hyper-surface, the 5D field equation $R_{4\alpha} = 0$ in the form (4) shows that $P_\alpha^\beta$ is exactly conserved: $P_{\alpha;\beta}^\beta = 0$. This implies that the physical quantities encoded by $P_\alpha^\beta$ act like a set of 4-currents. This view is supported by evaluating $P_\alpha^\beta$ for the canonical gauge. For this gauge, studies have shown that when the gauge is perturbed (e.g. by a shift in $x^4 = l$) there are connected changes in the value of the cosmological 'constant' of the spacetime and the effective mass of a particle which inhabits it [8]. This subject is still under investigation, but it is clear that the self-consistent definition of the mass of a test particle in a 5D manifold is difficult [9]. The problem is compounded by the need to make contact with recent advances in particle physics and data from the Large Hadron Collider, which support the view that particles acquire mass by interacting with a scalar field in the process known as the Higgs mechanism [10]. Hopefully, the Higgs mechanism may prove to be the quantum counterpart of the classical approach to the origin of mass known as Mach's Principle [11]. In Space-Time-Matter theory, as based on Campbell's theorem, matter arises mainly form the dependence of the metric coefficients on the extra coordinate, as is obvious from the $T_{\alpha\beta}$ of (3). This same mechanism is present in the $P_\alpha^\beta$ of (4). A logical interpretation of this tensor is therefore in terms of matter currents. And a logical definition for the rest mass of a particle is in terms of its associated scalar, via $mc/h \sim |P|$. The Appendix examines different physical interpretations of $P^{\alpha\beta}$. The simplest is $P^{\alpha\beta} = m u^\alpha u^\beta$, and leads to the standard (geodesic) equations of motion, as well as the inference that $m \sim |P|$.

The present work essentially outlines a class of solutions of the 5D field equations which follows from adopting the gauge (6). The 10 Einstein-like equations are satisfied by the energy-momentum tensor (7), which has contributions from the scalar field and an effective cosmological constant or vacuum term. The 4 conservation-type relations are satisfied by (8), whose scalar is proportional to the rest mass of the test particle as measured by the Compton wavelength $L = h/mc$. This identification fits with the last field equation ($R_{44} = 0$), which takes the form of the Klein-Gordon equation (9.1) when the extra dimension is timelike. Explicit solutions of the latter equation are given in standard texts on wave mechanics, but the present work holds for any of them. Generically, whether the extra dimension is timelike or spacelike, the effective cosmological constant and the 4D curvature scalar (9.2) are proportional to $1/L^2$ or $m^2$. In application to quantum mechanics, the picture is therefore of a particle whose Compton wavelength conforms to the curvature. This picture should be useful in studying the properties of particles in a vacuum-dominated universe.

Acknowledgements

The above grew out of earlier work by several members of the Space-Time-Matter group (http://astro.uwaterloo.ca/~wesson).

Appendix:  Physical consequences of $P^{\alpha\beta}{}_{;B} = 0$

The object of what follows is to work out the consequences of the 5D conservation relation $P^{\alpha\beta}{}_{;\beta} = 0$, where $P^{\alpha\beta}$ is a symmetric 4-tensor depending on the rest mass $m$ of a test particle and its 4-velocity $u^\alpha \equiv dx^\alpha / ds$.  The metric tensor $g_{\alpha\beta}$ of the 4D part of the 5D metric is used to raise and lower indices.  As in the main text, the 4D covariant derivative is denoted by a semicolon and the 4D partial derivative by a comma.  Three different forms for $P^{\alpha\beta}$ will be considered, to see what equations of motion they entail.

The simplest form is

$$P^{\alpha\beta} = m u^\alpha u^\beta \quad . \tag{1.1}$$

This is obviously symmetric, and matches the physical dimensions of $P^{\alpha\beta}\left(= L^{-1}\right)$ if atomic units are used to measure.  This means in practice that the particle's mass is geometrized by the Compton wavelength $h / mc$, where however the physical constants will be absorbed and dimensionless numerical factors ignored for algebraic ease.  Commonly, the 4-velocities are normalized by setting their product to ±1 (depending on the signature).  The resulting relation can be differentiated, and when the indices are rearranged provides a convenient starting point for the calculation:

$$u^\alpha u_\alpha = \varepsilon = \pm 1 \;, \quad u^\alpha{}_{;\beta} u_\alpha = 0 \quad . \tag{1.2}$$

It is also convenient to write (1.1) as $P^{\alpha\beta} = (mu^\beta)u^\alpha$, which differentiated and set to zero yields

$$(mu^\beta)_{;\beta} u^\alpha + mu^\beta u^\alpha_{;\beta} = 0 \quad . \tag{1.3}$$

Contracting this with $u_\alpha$ and using the first member of (1.2) gives

$$\varepsilon(mu^\beta)_{;\beta} + mu^\beta (u^\alpha_{;\beta} u_\alpha) = 0 \quad . \tag{1.4}$$

The last term in parentheses here is however zero by the second member of (1.2), so

$$(mu^\beta)_{;\beta} = p^\beta_{;\beta} = 0 \quad . \tag{1.5}$$

This relation is sometimes called the continuity equation for the momenta. It is identical in form to the continuity equation for a dust cloud, when the particle mass is replaced by the fluid density. Substituting (1.5) into (1.3) yields

$$u^\alpha_{;\beta} u^\beta = 0 \quad . \tag{1.6}$$

This is the shorthand version of the geodesic equation, which can also be obtained by extremizing the interval via $\delta\left[\int ds\right] = 0$ and reads

$$\frac{du^\alpha}{ds} = \Gamma^\alpha_{\gamma\delta} u^\gamma u^\delta = 0 \quad . \tag{1.7}$$

We see that when $P^{\alpha\beta}$ has the form (1.1), the four equations of motion (1.6) / (1.7) are independent of $m$.

A more complicated form of $P^{\alpha\beta}$ is

$$P^{\alpha\beta} = mu^{\alpha}u^{\beta} + kmg^{\alpha\beta} \quad . \tag{2.1}$$

Here $k$ is a dimensionless constant. The second term in (2.1) is allowable, of course, because the metric tensor behaves like a constant under covariant differentiation (see below). Following the same procedure as in the preceding paragraph, the analog of (1.5) is found to be

$$\varepsilon\left(mu^{\beta}\right)_{;\beta} + km_{,\beta}u^{\beta} = 0 \quad . \tag{2.2}$$

This implies that the momenta no longer have zero divergence along the particle's path, since

$$p^{\beta}_{;\beta} = -(k/\varepsilon)(dm/ds) \quad . \tag{2.3}$$

Putting (2.2) / (2.3) back into previous relations, the analog of (1.6) is found to be

$$u^{\alpha}_{;\beta}u^{\beta} = k\left[\frac{u^{\alpha}u^{\beta}}{\varepsilon} - g^{\alpha\beta}\right]\left(\frac{m_{,\beta}}{m}\right) \quad . \tag{2.4}$$

The term in brackets here is formally the same as the projector of general relativity, which acting on a 4-vector eliminates the component parallel to the 4-velocity and preserves the component in the orthogonal direction. Physically, the $k$-term in (2.1) is analogous to the cosmological-constant term $\Lambda g_{\alpha\beta}$ in Einstein's equations, and is likewise expected to be relatively small.

A third form of $P^{\alpha\beta}$ which is especially relevant to particles travelling on null 5D geodesics is given by taking the simple expression (1.1) but *not* constraining the 4-velocities by the normalization condition (1.2), thus:

$$P^{\alpha\beta} = mu^\alpha u^\beta , \quad u^\alpha u_\alpha \neq \text{constant} . \tag{3.1}$$

It will be seen below that foregoing the normalization condition leads to equations of motion which involve the velocity in the extra dimension $\left(u^4 \equiv dx^4/ds\right)$. This is relevant, because while some workers in space-time-matter theory have sought to identify $m$ in terms of $x^4$, others in membrane theory have sought to identify it in terms of $dx^4/ds$, where $s$ is the 4D proper time. It is perforce necessary to use the latter as parameter, rather than the 5D interval $S$, when the path is 5D null. In fact, it has become popular to take the view that *all* particles, even massive ones, travel on null paths with $dS^2 = 0$ in 5D, since this allows both timelike and null paths with $ds^2 \geq 0$ in 4D. In addition, it should be noted that it may be too physically restrictive to impose the condition $\left|u^\alpha u_\alpha\right| = 1$ of (1.2) as well as the condition $dS^2 = 0$. To Illustrate this, consider the 5D Minkowski metric $dS^2 = dt^2 - \left(dx^2 + dy^2 + dz^2\right) - dl^2$. Dividing this by $ds^2$, and imposing both noted conditions, leads to $\left|dl/ds\right| = 1$, which is very restrictive. Therefore, only the 5D null condition is adopted. The derivative of this gives

$$\left(dS^2/ds^2\right) = 0 = u^\alpha u_\alpha + \varepsilon u^4 u_4$$

$$u^\alpha_{;\beta} u_\alpha = -\varepsilon u^4_{;\beta} u_4 . \tag{3.2}$$

The second member of this is the analog of the second member of (1.2). Forming $P^{\alpha\beta}_{;\beta}=0$ from (3.1) leads to

$$\left(mu^{\beta}\right)_{;\beta}u^{\alpha}+mu^{\beta}u^{\alpha}_{;\beta}=0 \quad . \tag{3.3}$$

Contracting this with $u_\alpha$ and employing both members of (3.2) yields

$$\left(mu^{\beta}\right)_{;\beta}u^{4}u_{4}+mu^{\beta}u^{4}_{;\beta}u_{4}=0 \quad . \tag{3.4}$$

Multiplying (3.3) by $u^4 u_4$ and (3.4) by $u^\alpha$, and taking the difference, gives

$$u^{\beta}u_{4}\left(u^{\alpha}_{;\beta}u^{4}-u^{4}_{;\beta}u^{\alpha}\right)=0 \quad . \tag{3.5}$$

Rearranging this results finally in

$$u^{\alpha}_{;\beta}u^{\beta}=\frac{u^{\alpha}}{u^{4}}\frac{du^{4}}{ds} \quad . \tag{3.6}$$

This is the analog of (1.6) and represents geodesic motion perturbed by a term that depends on the acceleration in the extra dimension.